\documentclass[journal,twocolumn,10pt,twoside]{IEEEtran}
\normalsize

\ifCLASSINFOpdf
\else
\fi

\usepackage{amsmath}
\usepackage{amssymb}
\usepackage{cite}
\usepackage{graphicx}
\usepackage{algorithm}
\usepackage{algpseudocode}
\usepackage{subfigure}
\usepackage{multirow} 
\usepackage{longtable}
\usepackage{threeparttable}
\usepackage{caption3}
\usepackage{array}
\usepackage{cases}
\usepackage{color}
\usepackage[T1]{fontenc}
\usepackage[utf8]{inputenc}
\usepackage{authblk}
\usepackage{subfigmat}

\newcommand\mycom[2]{\genfrac{}{}{0pt}{}{#1}{#2}}

\begin{document}
\title{A Low-Complexity Adaptive Multisine Waveform Design for Wireless Power Transfer}
\author{Bruno Clerckx and Ekaterina Bayguzina
\thanks{The authors (email: \{b.clerckx,ekaterina.bayguzina08\}@imperial.ac.uk) are with Imperial College London, UK. The work has been partially supported by the EPSRC under grants EP/K502856/1, EP/L504786/1 and EP/P003885/1. }}
\maketitle

\begin{abstract} Far-field Wireless Power Transfer (WPT) has attracted significant attention in the last decade. Recently, channel-adaptive waveforms have been shown to significantly increase the DC power level at the output of the rectifier. However the design of those waveforms is generally computationally complex and does not lend itself easily to practical implementation. We here propose a low-complexity channel-adaptive multisine waveform design whose performance is very close to that of the optimal design. Performance evaluations confirm the benefits of the new design in various rectifier topologies.
\end{abstract}

\begin{IEEEkeywords} Wireless Power Transfer, Waveform, Rectenna
\end{IEEEkeywords}

\IEEEpeerreviewmaketitle

\vspace{-0.5cm}
\section{Introduction}


\IEEEPARstart{W}{ireless} Power Transfer (WPT) via radio-frequency radiation is nowadays attracting more and more attention, with clear applications in Wireless Sensor Networks (WSN) and Internet of Things (IoT) \cite{Visser:2013}. The major challenge facing far-field wireless power designers is to find ways to enhance the end-to-end power transfer
efficiency, or equivalently increase the DC power level at the output of the rectenna
without increasing the transmit power, and for devices located tens to hundreds of meters away from the transmitter. To that end, the vast majority of the technical efforts in the literature has been devoted to the design of efficient rectennas, a.o. \cite{Visser:2013}, so as to increase the RF-to-DC conversion efficiency. 
\par Another promising approach is to design efficient WPT signals (including waveforms, beamforming, power allocation) \cite{Zeng:2016,Clerckx:2016b}. Indeed, the transmit signal design has a major impact on the end-to-end power transfer efficiency as it influences the signal strength at the input of the rectenna but also the RF-to-DC conversion efficiency due to the rectifier nonlinearity. Traditional approaches consist in using waveforms that exhibit high PAPR \cite{Trotter:2009,OptBehaviour,Valenta:2015,Bolos:2016}. Unfortunately, those approaches were heuristic and ignored the presence of the time-varying wireless propagation channel that is subject to multipath and fading. Multipath has for consequence that the transmit and the received (at the input of the rectenna) waveforms are completely different. Hence, the transmit waveforms should be designed in accordance with the channel status. However, all those approaches in the RF literature \cite{Trotter:2009,OptBehaviour,Valenta:2015,Bolos:2016} are based on an open-loop architecture with waveforms being static. A systematic design and optimization of channel-adaptive waveform and signal for WPT has recently been tackled in \cite{Clerckx:2016b,Clerckx:2015} and further extended in \cite{Huang:2016} for large-scale WPT. Gains over various baseline waveforms have been shown to be very significant. This adaptive design leads to a closed-loop architecture where the channel state information is acquired to the transmitter and the
transmit signal is dynamically adapted so as to maximize the DC power at the output of the rectifier, accounting for the wireless channel and the rectifier nonlinearity. Contrary to what
is claimed in \cite{Trotter:2009,OptBehaviour,Valenta:2015,Bolos:2016}, maximizing PAPR was shown in \cite{Clerckx:2016b} not to be a right approach to design efficient WPT signal. High PAPR signals are useful if the channel is frequency flat, not in the presence of multipath and frequency selectivity.
\par Unfortunately, those optimized and adaptive waveforms do not lend themselves easily to practical implementation because they result from a non-convex optimization problem that would require to be solved real-time as a function of the channel state information (CSI). This is computationally intensive. Those waveforms can therefore be viewed as benchmarks we should aim for performance-wise. What we propose in this paper is a design of multi-sine waveform, adaptive to CSI, whose performance is very close to the optimal design of \cite{Clerckx:2015,Clerckx:2016b} but whose complexity is significantly lower. Indeed the proposed design results from a simple scaled matched filter that has the effect of allocating more (resp.\ less) power to the frequency components corresponding to large (resp.\ weak) channel gains.   
  
\par \textit{Notations:} Bold letters stand for vectors or matrices. $|.|$ and $\left\|.\right\|$ refer to the absolute value of a scalar and the 2-norm of a vector. $\mathcal{E}\left\{.\right\}$ refers to the averaging operator. 
\vspace{-0.2cm}
\section{WPT System Model}\label{system_model}
\vspace{-0.0cm}
\subsection{Transmit and Received Signal}
Consider a multisine signal $x(t)=\Re\big\{\sum_{n=0}^{N-1}w_{n}e^{j2\pi f_n t}\big\}$ (with $N$ sinewaves) transmitted at time $t$ over a single antenna
with $w_{n}=s_{n}e^{j\phi_{n}}$ where $s_{n}$ and $\phi_{n}$ refer to the amplitude and phase of the $n^{th}$ sinewave at frequency $f_n$, respectively. We assume that the frequencies are evenly spaced, i.e.\ $f_n=f_0+n\Delta_f$ with $\Delta_f$ the frequency spacing. The magnitudes and phases of the sinewaves are collected into vectors $\mathbf{s}$ and $\mathbf{\Phi}$, whose  $n^{th}$ entry writes as $s_{n}$ and $\phi_{n}$, respectively. The transmit power constraint is given by $\mathcal{E}\big\{\left|x\right|^2\big\}=\frac{1}{2}\left\|\mathbf{s}\right\|^2\leq P$.

\par The transmitted sinewaves propagate through a multipath channel, characterized by $L$ paths whose delay, amplitude and phase are respectively denoted as $\tau_l$, $\alpha_l$, $\xi_l$, $l=1,\ldots,L$. The signal received at the single-antenna receiver is written as
\begin{align}
y(t)
=\!\sum_{n=0}^{N\!-\!1}s_{n}A_{n} \cos(\varpi_n t\!+\!\psi_{n})=\!\Re\left\{\sum_{n=0}^{N\!-\!1} h_n w_n e^{j \varpi_n t}\right\}\label{received_signal_ant_m}
\end{align}
where $h_{n}=A_{n}e^{j \bar{\psi}_{n}}=\sum_{l=0}^{L-1}\alpha_l e^{j(-2\pi f_n\tau_l+\xi_{l})}$ is the channel frequency response at frequency $f_n$ and $\varpi_n=2\pi f_n$. $A_{n}$ and $\psi_{n}$ are defined such that $A_{n}e^{j \psi_{n}}\!=\!A_{n}e^{j \left(\phi_{n}+\bar{\psi}_{n}\right)}\!=\!e^{j \phi_{n}}h_{n}$.
 
\vspace{-0.2cm}
\subsection{Antenna Equivalent Circuit Model}
\begin{figure}
 \begin{minipage}[c]{.5\linewidth}
   \centerline{\includegraphics[width=0.9\columnwidth]{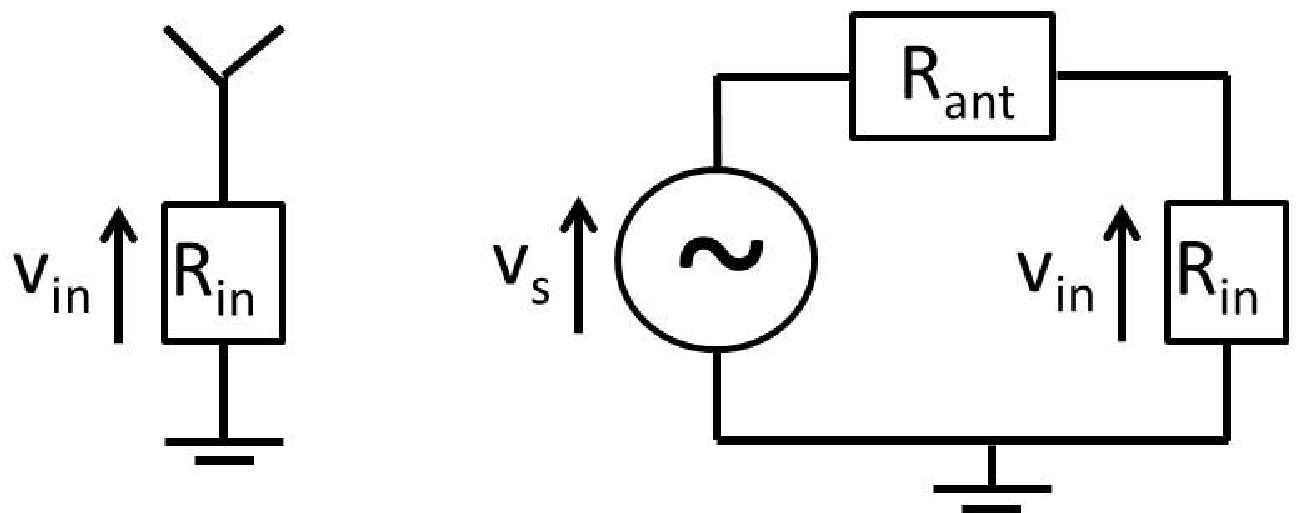}}
  \end{minipage}\hfill
 \begin{minipage}[c]{.5\linewidth}
   \centerline{\includegraphics[width=0.9\columnwidth]{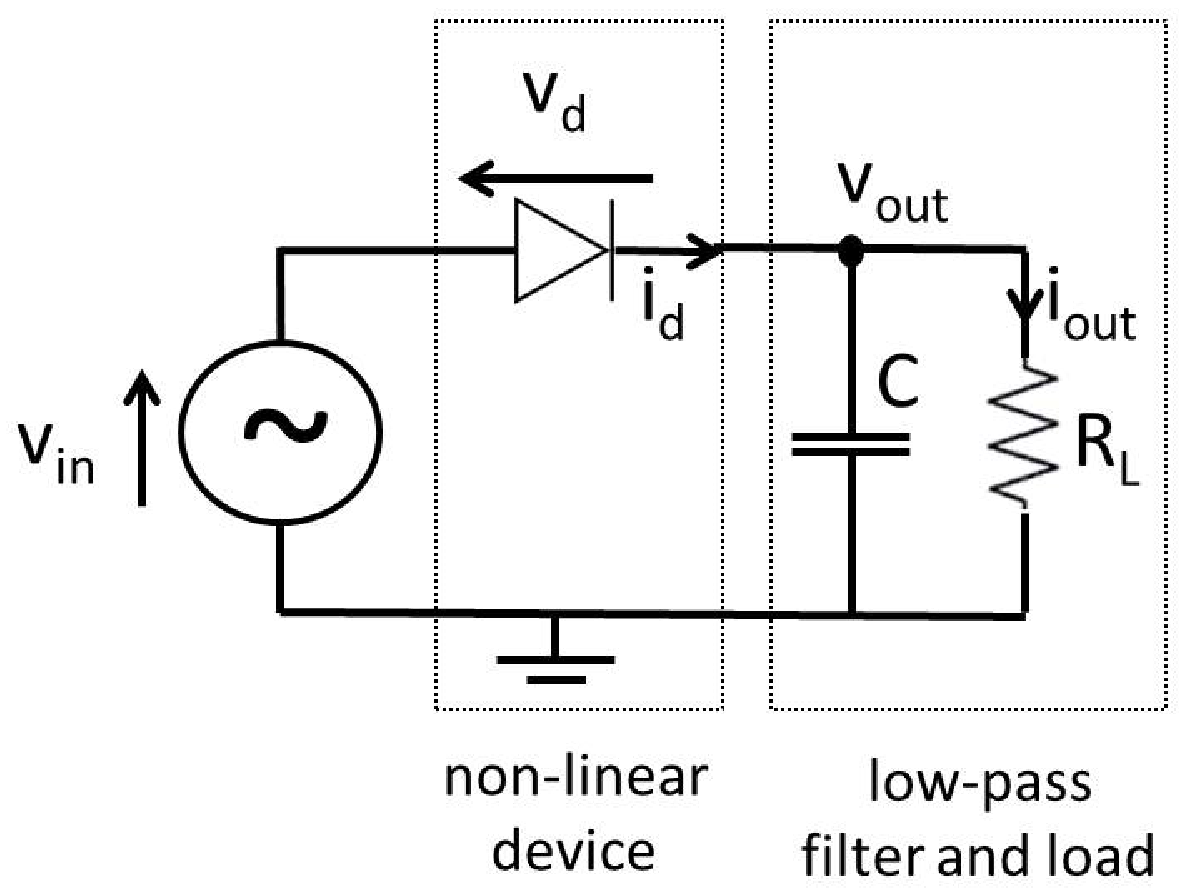}}
  \end{minipage}
  \caption{Antenna equivalent circuit (left) and a single diode rectifier (right).}
  \label{antenna_model}
  \vspace{-0.2cm}
\end{figure}

The antenna model reflects the power transfer from the
antenna to the rectifier through the matching network. As illustrated in Fig \ref{antenna_model}(left), a lossless antenna
can be modelled as a voltage source $v_s(t)$ followed by a
series resistance $R_{ant}$. Let $Z_{in} = R_{in} + j X_{in}$ denote the
input impedance of the rectifier with the matching network.
Assuming perfect matching ($R_{in} = R_{ant}$, $X_{in} = 0$), all the
available RF power $P_{in,av}$ is transferred to the rectifier and
absorbed by $R_{in}$, so that $P_{in,av} = \mathcal{E}\big\{\left|v_{in}(t)\right|^2\big\}/R_{in}$ and $v_{in}(t)=v_{s}(t)/2$. Since $P_{in,av} =\mathcal{E}\big\{\left|y(t)\right|^2\big\}$, $v_{in}(t)=y(t)\sqrt{R_{in}}=y(t)\sqrt{R_{ant}}$. 

\vspace{-0.3cm}
\subsection{Rectifier and Diode Non-Linear Model}
Consider a rectifier composed of a single series diode followed by a low-pass filter with load as in Fig \ref{antenna_model}(right). Denoting the voltage drop across the diode as $v_d(t)=v_{in}(t)-v_{out}(t)$ where $v_{in}(t)$ is the input voltage to the diode and $v_{out}(t)$ is the output voltage across the load resistor, a tractable behavioural diode model is obtained by Taylor series expansion of the diode characteristic equation $i_d(t)=i_s \big(e^{\frac{v_d(t)}{n v_t}}-1 \big)$ (with $i_s$ the reverse bias saturation current, $v_t$ the thermal voltage, $n$ the ideality factor assumed equal to $1.05$) around a quiescent operating point $v_d=a$, namely $i_d(t)=\sum_{i=0}^{\infty }k_i' \left(v_d(t)-a\right)^i$
where $k_0'=i_s\big(e^{\frac{a}{n v_t}}-1\big)$ and $k_i'=i_s\frac{e^{\frac{a}{n v_t}}}{i!\left(n v_t\right)^i}$, $i=1,\ldots,\infty$.
Assume a steady-state response and an ideal low pass filter such that $v_{out}(t)$ is at constant DC level. Choosing $a=\mathcal{E} \left\{ v_d(t) \right\}=-v_{out}$, we can write $i_d(t)=\sum_{i=0}^{\infty }k_i' v_{in}(t)^i=\sum_{i=0}^{\infty }k_i' R_{ant}^{i/2} y(t)^i$.
Truncating the expansion to order 4, the DC component of $i_{d}(t)$ is the time average of the diode current, and is obtained as $i_{out}\approx k_0'+k_2' R_{ant}\mathcal{E}\left\{y(t)^2\right\}+k_4' R_{ant}^2\mathcal{E}\left\{y(t)^4\right\}$. 

\vspace{-0.2cm}
\section{A Low-Complexity Adaptive Waveform Design}\label{low_complexity}
Assuming the CSI (in the form of frequency response $h_{n}$) is known to the transmitter\footnote{Inspired by communication systems, we
could envision a pilot transmission phase (on the uplink) and a channel estimator at the power base station. Alternatively, approaches relying on
CSI feedback could be exploited \cite{Zeng:2016}.}, we aim at finding the set of amplitudes and phases $\mathbf{s},\mathbf{\Phi}$ that maximizes $i_{out}$. Following \cite{Clerckx:2016b}, this is equivalent to maximizing the quantity 
\begin{equation}\label{diode_model_2}
z_{DC}\left(\mathbf{s},\mathbf{\Phi}\right)=k_2 R_{ant}\mathcal{E}\left\{y(t)^2\right\}+k_4 R_{ant}^2\mathcal{E}\left\{y(t)^4\right\}
\end{equation}
where $k_i=\frac{i_s}{i!\left(n v_t\right)^i}$, $i=2,4$. Assuming $i_s=5 \mu A$, a diode ideality factor $n=1.05$ and $v_t=25.86 mV$, typical values are given by $k_2=0.0034$ and $k_4=0.3829$. 

\par The waveform design problem can therefore be written as
\begin{equation}\label{P2}
\max_{\mathbf{s},\mathbf{\Phi}} \hspace{0.2cm} z_{DC}(\mathbf{s},\mathbf{\Phi}) \hspace{0.3cm} \textnormal{subject to} \hspace{0.3cm} \frac{1}{2}\left\|\mathbf{s}\right\|_F^2\leq P,
\end{equation}
where $z_{DC}$ is given in \eqref{y_DC} after plugging $y(t)$ of \eqref{received_signal_ant_m} into \eqref{diode_model_2}.
\begin{table*}
\begin{align}\label{y_DC}
z_{DC}(\mathbf{s},\mathbf{\Phi})&=\frac{k_{2}}{2}R_{ant}\left[\sum_{n=0}^{N-1} s_{n}^2A_{n}^2\right]+\frac{3k_{4}}{8}R_{ant}^2\left[\sum_{\mycom{n_0,n_1,n_2,n_3}{n_0+n_1=n_2+n_3}}\Bigg[\prod_{j=0}^3s_{n_j}A_{n_j}\Bigg]\cos(\psi_{n_0}+\psi_{n_1}-\psi_{n_2}-\psi_{n_3})\right].
\end{align}\hrulefill
\end{table*}

\par From \cite{Clerckx:2015,Clerckx:2016b}, the optimal phases are given by $\phi_{n}^{\star}=-\bar{\psi}_{n}$ while the optimum amplitudes result from a non-convex posynomial maximization problem which can be recasted as a Reversed Geometric Program (GP) and solved iteratively but does not easily lend itself to practical implementation. Indeed, according to \cite{Chiang:2005}, Reversed GP takes exponential time to compute an optimal solution. Interestingly, as noted in \cite{Clerckx:2016b}, the optimized waveform has a tendency to allocate more power to frequencies exhibiting larger channel gains. Motivated by this observation, we propose here, as a suboptimal solution to \eqref{P2}, a simple, closed-form and low-complexity strategy, denoted as scaled matched filter (SMF), that selects the phases as $\phi_{n}^{\star}$ but chooses the amplitudes of sinewaves according to\footnote{Note that, following \cite{Clerckx:2016b}, the SMF strategy is easily extendable to multiple transmit antennas by replacing $A_n$ with the norm of the vector channel.}
\begin{equation}\label{SMF}
s_n=c A_n^{\beta}
\end{equation}
where $c$ is the constant satisfying the transmit power constraint $\frac{1}{2}\sum_{n=0}^{N-1}s_n^2=P$ and $\beta\geq 1$. The complex weight of the SMF waveform  on frequency $n$ is finally given in closed form as
\begin{equation}
w_n=e^{-j \bar{\psi}_{n}} A_n^{\beta} \sqrt{\frac{2P}{\sum_{n=0}^{N-1}A_n^{2\beta}}}.\label{scaling_MF}
\end{equation}
The SMF waveform design is only a function of a single parameter, namely $\beta$.
By taking $\beta=1$, we get a matched filter-like behaviour, where the amplitude of sinewave $n$ is linearly proportional to $A_n$. This is reminiscent of maximum ratio combining (MRC) in communication. On the other hand, by scaling $A_n$ using an exponent $\beta>1$, we further amplify the strong frequency components and attenuate the weak ones.

\par $\beta$ can either be optimized on a channel basis or be fixed once for all. Plugging \eqref{scaling_MF} into \eqref{y_DC}, we get \eqref{z_DC_SMF}. For a given channel realization, the best $\beta$ can be obtained as the solution of the unconstrained optimization problem $\beta^{\star}=\arg\max_{\beta} \hspace{0.1cm} z_{DC,SMF}$, which can be solved numerically using Newton's method.

\begin{table*}
\begin{equation}\label{z_DC_SMF}
z_{DC,SMF}=k_2 R_{ant} P \left[\sum_{n=0}^{N-1}\frac{A_n^{2(\beta+1)}}{\sum_{n=0}^{N-1}A_n^{2\beta}}\right]+\frac{3k_{4}}{2}k_4 R_{ant}^2 P^2\left[\sum_{\mycom{n_0,n_1,n_2,n_3}{n_0+n_1=n_2+n_3}}\frac{\prod_{j=0}^3 A_{n_j}^{\beta+1}}{\left[\sum_{n=0}^{N-1}A_n^{2\beta}\right]^2}\right]
\end{equation}
\hrulefill
\end{table*}

\par
\begin{figure}
\centerline{\includegraphics[width=\columnwidth]{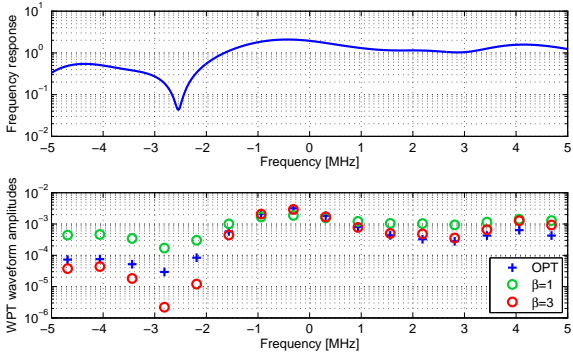}}
  \caption{Frequency response of the wireless channel and WPT waveform magnitudes ($N=16$) for 10 MHz bandwidth.}
  \label{Freq_response_channel1}
  \vspace{-0.2cm}
\end{figure}

 In order to get some insights into the SMF strategy \eqref{SMF}, we consider a frequency selective channel whose frequency response is given by Fig \ref{Freq_response_channel1} (top), a transmit power of -20dBm, $N=16$ sinewaves centered around 5.18GHz with a frequency gap fixed as $\Delta_f=B/N$ and $B=10$MHz. Assuming such a channel realization, we compare in Fig \ref{Freq_response_channel1} (bottom) the magnitudes of the SMF waveform (with $\beta=1,3$) and of the optimum (OPT) waveform obtained using the Reversed GP algorithm derived in \cite{Clerckx:2015,Clerckx:2016b}. The OPT waveform has a tendency to allocate more power to frequencies exhibiting larger channel gains. Choosing $\beta=1$ would allocate power proportionally to the channel strength but has a tendency to underestimate the power to be allocated to strong channels and overestimate the power to be allocated to weak channels. On the other hand, suitably choosing $\beta>1$ better emphasizes the strong channels and de-emphasizes the weak channels. 

\section{Performance Evaluations}\label{eval}
\par
\begin{figure}
\centerline{\includegraphics[width=0.9\columnwidth]{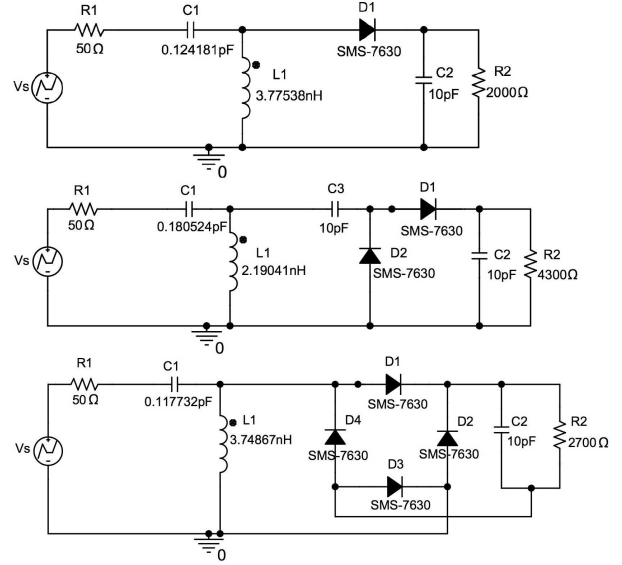}}
  \caption{Single series, voltage doubler and diode bridge rectifiers.}
  \label{rectenna_circuit}
  \vspace{-0.2cm}
\end{figure}

 In this section, we evaluate the performance of the waveforms using the rectifier configurations of Fig \ref{rectenna_circuit}. We consider a point-to-point scenario representative of a WiFi deployment at a center frequency of 5.18GHz with a 36dBm transmit power, isotropic transmit antennas, 2dBi receive antenna gain and 58dB path loss in a large open space environment with a NLOS channel power delay profile obtained from model B \cite{Medbo:1998b}. Taps are modeled as i.i.d.\ circularly symmetric complex Gaussian random variables and normalized such that the average received power is $P_{in,av}=-20$dBm. The $N$ sinewaves are centered around 5.18GHz with $\Delta_f=B/N$ and $B=10$MHz.
\par The rectennas in Fig \ref{rectenna_circuit} are optimized for a multisine input signal composed of 4 sinewaves\footnote{Due to the channel frequency selectivity, out of a
large number of sinewaves, only a few are allocated significant
power (e.g. as in Fig \ref{Freq_response_channel1}).} and the available RF power of $-20$dBm. The package parasitics of components are ignored in all simulator models. The L-matching network is optimized together with the load resistance using ADS with the objective to maximize the output DC power and minimize impedance mismatch due to a signal of varying instantaneous power.    

\par
\begin{figure*}
\begin{subfigmatrix}{2}
\subfigure[Average $z_{DC}$.]{\label{a}\includegraphics[width = 9cm]{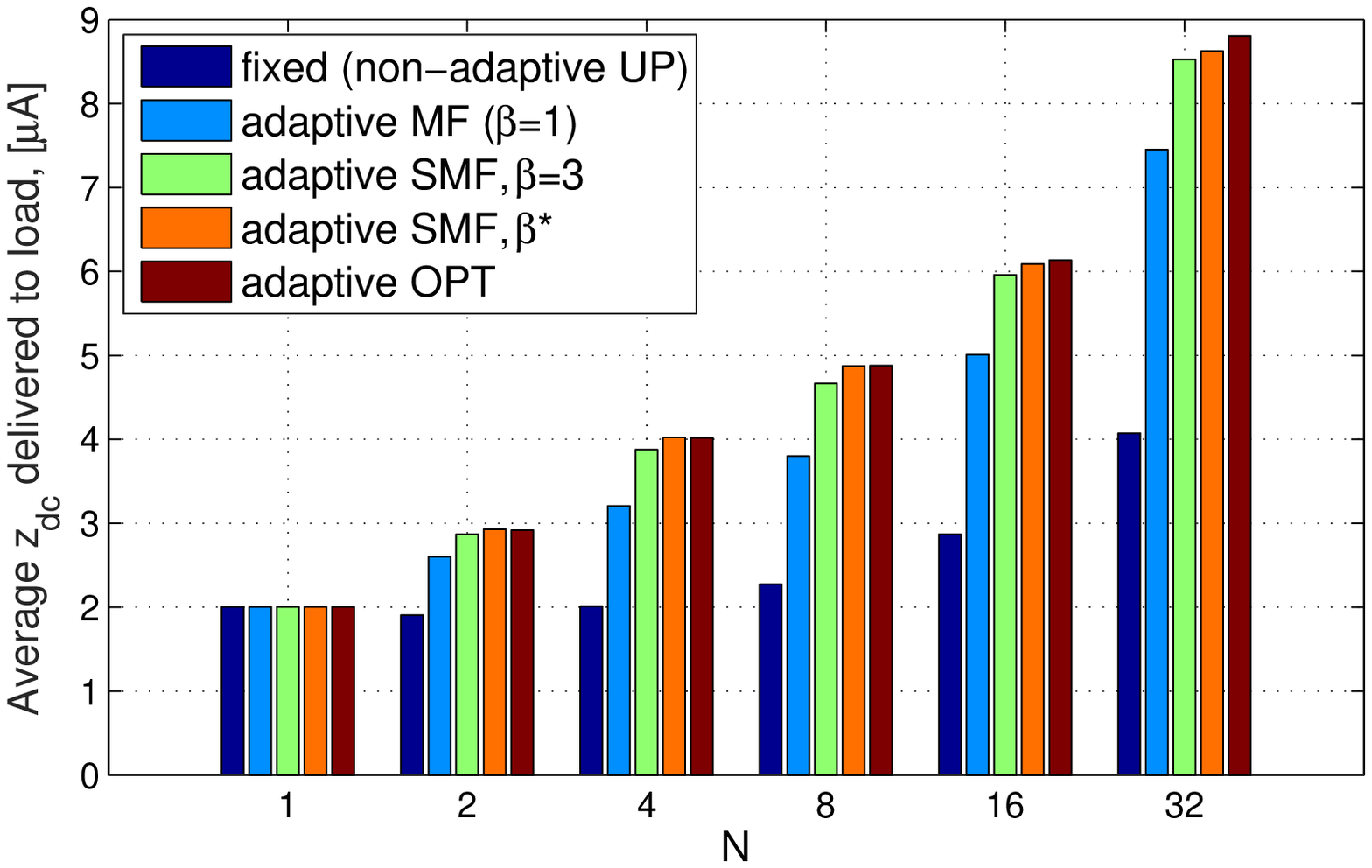}}
\subfigure[Single series.]{\label{b}\includegraphics[width = 9cm]{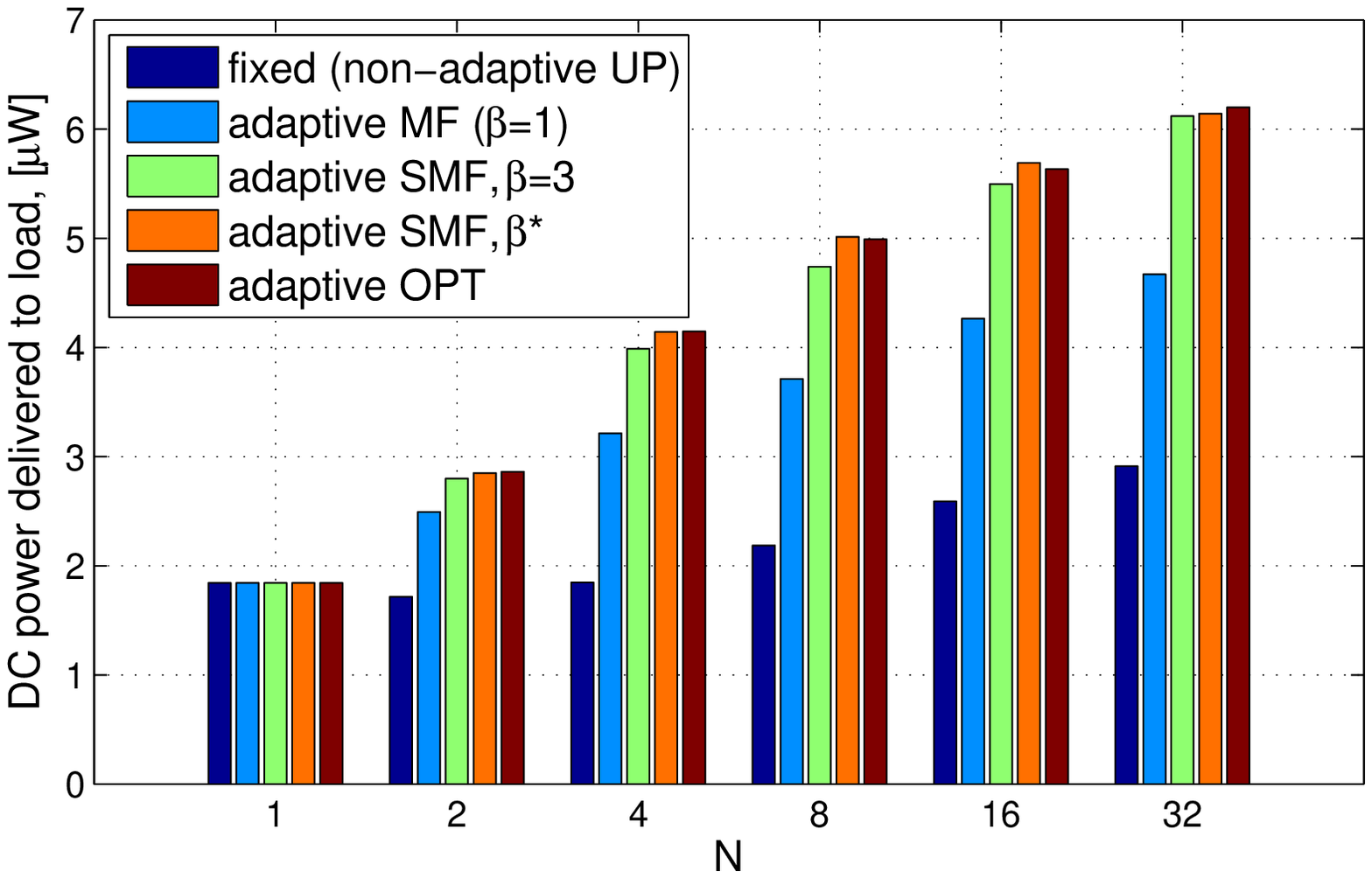}}
\vspace{-0.25cm}
\subfigure[Voltage doubler.]{\label{b}\includegraphics[width = 9cm]{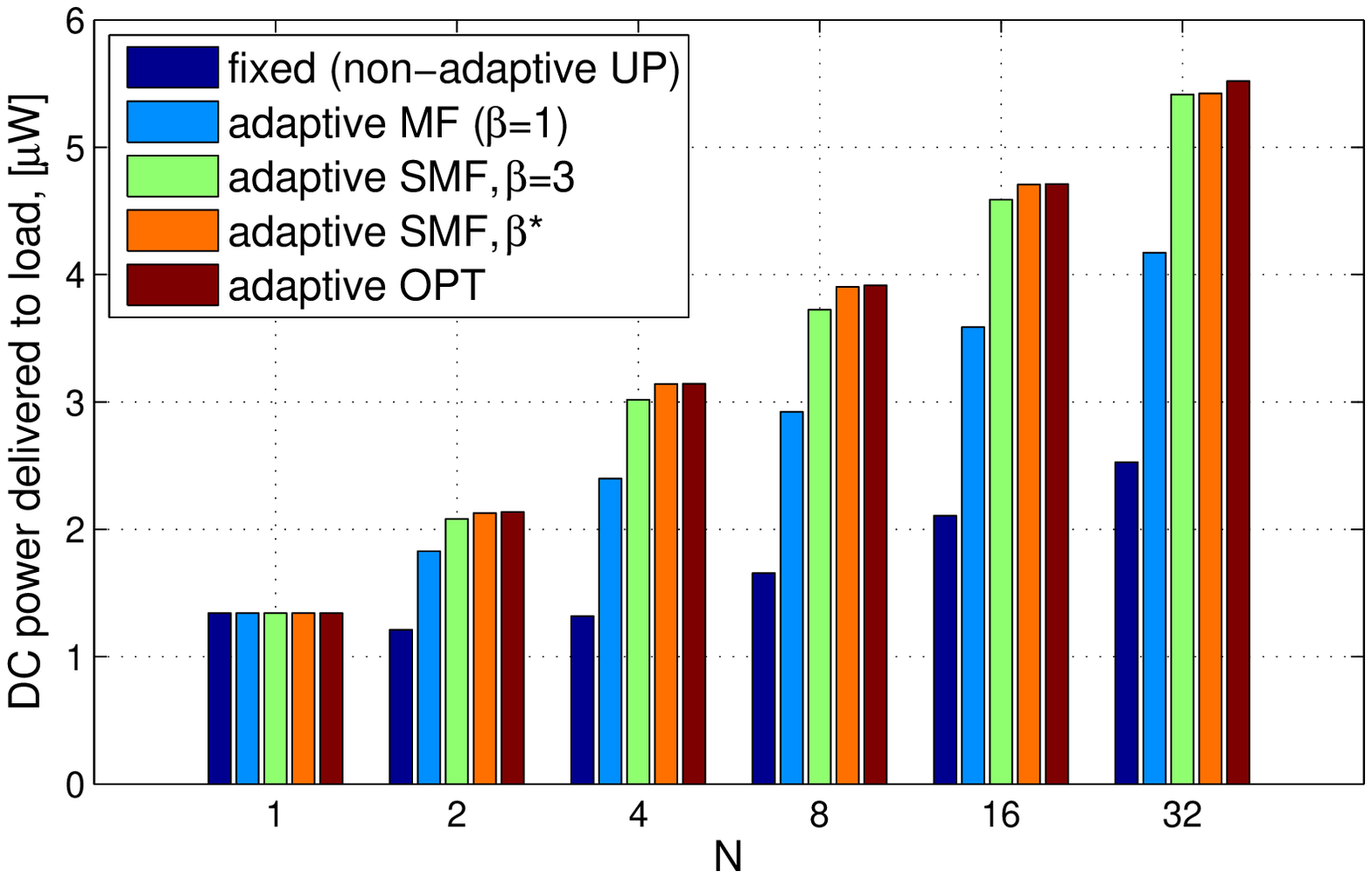}}
\subfigure[Diode bridge.]{\label{b}\includegraphics[width = 9cm]{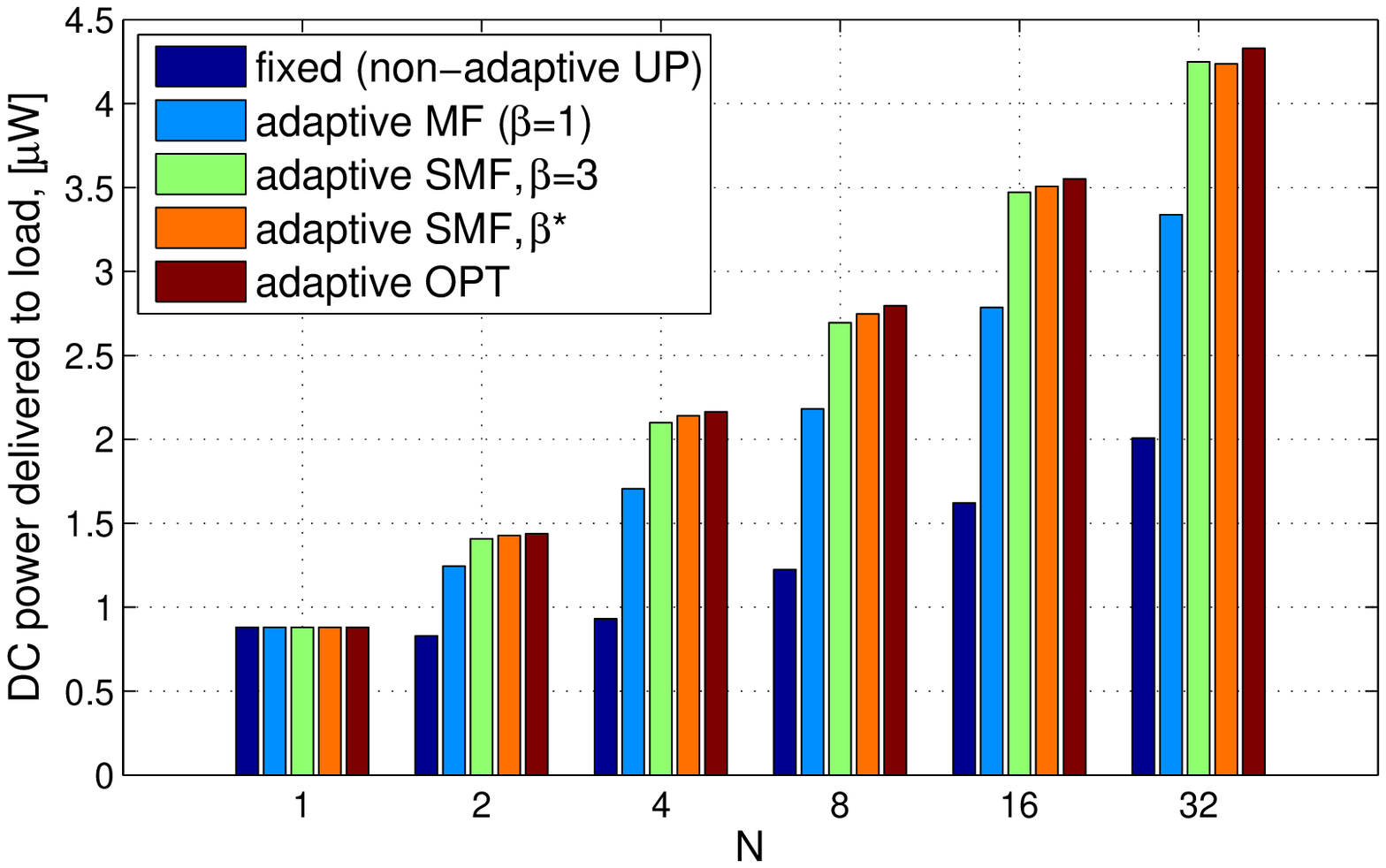}}
\end{subfigmatrix}
\caption{Average $z_{DC}$ and Average DC power delivered to the load as a function of $N$ for various rectifiers.}
\label{pspice_results}
\vspace{-0.4cm}
\end{figure*}

 In Fig \ref{pspice_results}(a), we display $z_{DC}$ averaged over many channel realizations for various waveforms. The traditional fixed waveform is not adaptive to CSI and is obtained by allocating power uniformly (UP) across sinewaves and fixing the phases $\phi_n$ as 0 \cite{Trotter:2009,OptBehaviour,Valenta:2015}. Adaptive MF is a particular case of the proposed SMF with $\beta=1$. SMF with $\beta^{\star}$ refers to the SMF waveform where $\beta$ is optimized on each channel realization using the Newton's method. Adaptive OPT is the optimal strategy resulting from the reversed GP algorithm derived in \cite{Clerckx:2015,Clerckx:2016b}. We note that the proposed waveform strategy SMF with $\beta=3$ comes very close to the optimal performance but incurs a significantly lower complexity since the weights are given in closed-form.

\par In Fig \ref{pspice_results}(b)(c)(d), we evaluate the waveform performance using PSpice simulations. To that end, the waveforms after the wireless channel have been used as inputs to the rectennas of Fig \ref{rectenna_circuit} and the DC power delivered to the load has been observed. The average DC power, where averaging is done over many realizations of the wireless channel, is displayed in Fig \ref{pspice_results}(b)(c)(d) as a function of $N$. We confirm the observations made using the $z_{DC}$ metric in Fig \ref{pspice_results}(a), namely that the performance of SMF with $\beta=3$ or $\beta^{\star}$ is very close to that of OPT despite the much lower design complexity. The PSpice evaluations also confirm the benefits of the SMF and OPT waveforms over the conventional non-adaptive UP multisine waveform and the usefulness of the waveform design methodology of \cite{Clerckx:2016b} in a wide range of rectifier configurations. Results also highlight the importance of efficient waveform design for WPT. Taking for instance Fig \ref{pspice_results}(b), we note that the RF-to-DC conversion efficiency jumps from less than 20\% to over 60\% by making use of 32 sinewaves rather than a single sinewave. We also note that at low average input power, a single series rectifier is preferable over the voltage doubler or diode bridge, which is inline with observations made in \cite{OptBehaviour}. 
\par Results also highlight that the non-linear rectifier model and the waveform design and optimization are valid for the multiple-diode rectenna configurations. This is because only a subset of diodes are conducting during a half-cycle of the input waveform and such operating point $a$ can be found so that $z_{DC}$ can still be expressed as in \eqref{diode_model_2}.
In case of the voltage doubler, ignoring the voltage gain due to the matching network and the forward voltage drops across the diodes, the capacitor $C_3$ is charged to the peak of the input waveform $\hat{v}_{in}$ during the negative half-cycle. 
During the positive half-cycle, the voltage across the diode $D_1$ can be obtained as $v_{d1}(t)\approx v_{in}(t)+\hat{v}_{in}-v_{out}(t)$, so the appropriate choice of the operating point is $a=\hat{v}_{in}-v_{out}$.
In case of the diode bridge, when the diodes $D_1$ and $D_3$ are conducting during the positive half-cycle, the current flowing through and the voltage across the diodes $D_1$ and $D_3$ are equal, so  $v_{d1}(t)=v_{in}(t)-v_{out}(t)-v_{d3}(t)$ and $v_{d1}(t)=v_{d3}(t)$. Since $\mathcal{E}\left\{2v_{d3}(t)\right\}=-v_{out}$, the operating point is calculated as $a=-\frac{v_{out}}{2}$, and $z_{DC}$ can be obtained for all the diodes as in \eqref{diode_model_2}. The choice of $a$ affects the values of $k_i'$, but the waveform optimization algorithm is only sensitive to $k_2$ and $k_4$ (and therefore not $a$) \cite{Clerckx:2016b}. Consequently, the solution to the optimization problem and the design of low-complexity waveforms remain unchanged for multiple-diode rectennas.
\par Last but not least, it is important to note that while the optimized and proposed waveforms are adaptive to the CSI, the rectifiers as designed and simulated are not adaptive to the CSI. This is because the wireless channel changes in practice so quickly that it would be impractical for energy-constrained devices to dynamically compute and adjust the matching and the load as a function of the channel. This adaptive signal approach makes the transmitter smarter and decreases the need for power-hungry optimization at the devices.

\vspace{-0.2cm}
\section{Conclusions}\label{conclusions}
The paper derives a methodology to design low-complexity adaptive waveforms for WPT. Assuming the CSI is available to the transmitter, the waveforms are designed such that more (resp. less) power is allocated to the frequency components corresponding to large (resp. weak) channel gains. They are shown through realistic simulations to achieve performance very close to the optimal waveforms in various rectenna configurations. Thanks to their low complexity, the proposed waveforms are very suitable for practical implementation.

\ifCLASSOPTIONcaptionsoff
  \newpage
\fi

\end{document}